# Diagnosis Prevalence vs. Efficacy in Machine-learning Based Diagnostic Decision Support

Gil Alon[1], Elizabeth Chen, PhD[1], Guergana Savova, PhD[2], Carsten Eickhoff, PhD[1]
[1]Center for Biomedical Informatics, Brown University, Providence, RI, United States
[2]Computational Health Informatics Program, Boston Children's Hospital and Harvard Medical School, Boston, MA, United States

**Abstract**

*Many recent studies use machine learning to predict a small number of ICD-9-CM codes. In practice, on the other hand, physicians have to consider a broader range of diagnoses. This study aims to put these previously incongruent evaluation settings on a more equal footing by predicting ICD-9-CM codes based on electronic health record properties and demonstrating the relationship between diagnosis prevalence and system performance. We extracted patient features from the MIMIC-III dataset for each admission. We trained and evaluated 43 different machine learning classifiers. Among this pool, the most successful classifier was a Multi-Layer Perceptron. In accordance with general machine learning expectation, we observed all classifiers' F1 scores to drop as disease prevalence decreased. Scores fell from 0.28 for the 50 most prevalent ICD-9-CM codes to 0.03 for the 1000 most prevalent ICD-9-CM codes. Statistical analyses showed a moderate positive correlation between disease prevalence and efficacy (0.5866).*

**Introduction**

Studies have focused on estimating the error rate of individual physician's diagnosis and the harms of diagnostic error. Barnett et al. (2019)[1] reported individual physician's diagnosis accuracy of 62.5% and noted increases to 85.6% if teams of nine physicians worked together in diagnosing the patient. Singh et al. (2014)[2] combined data from three prior studies to estimate the frequency of misdiagnosis, finding a 5% diagnostic error rate in those three datasets. When extrapolating those results to the entire United States (U.S.) population, Singh et al. estimate that about 1 in 20 adults may be misdiagnosed every year and half of these errors could be harmful. Shojania et al. (2003)[3] studied autopsy-identified misdiagnoses and estimated that in the U.S. between 8.4 and 24.4 percent of patients could experience a major error in their diagnosis.

Zwaan et al. (2010)[4] tried to determine the cause of diagnostic error using randomly selected patients from 21 hospitals in the Netherlands. The researchers concluded that human failure was one of the main causes of diagnostic error. Zwaan further found that the primary causes of diagnostic adverse events were knowledge-based mistakes (physicians did not have sufficient information available) and information transfer problems (doctors did not receive current updates about the patient). In 2014, the Controlled Risk Insurance Company (CRICO)[5], a division of the Risk Management Foundation of Harvard Medical Institutions Incorporated, studied the causes of diagnostic error. Using their own data, they found that 1% of malpractice claims involved an error of the patient failing to engage (recognize their symptoms and seek treatment) and that a large number of cases may have additionally suffered from diagnostic errors. Gandhi et al. (2006)[6] examined 181 closed malpractice claims. The researchers found that 79% of claims were involved failures in judgement of the practitioner, but also found that 46% of claims had patient-related factors. Their ultimate conclusion stated that diagnostic error is the breakdown of multiple- human and system- factors. Likewise, The National Academies of Sciences (2015)[7] concludes that there are numerous causes of diagnostic error and no single one can explain or cause a misdiagnosis. However, Zwaan further describes that diagnostic error contributes more heavily to the mortality rate of the patient than any other adverse event.

Therefore, the introduction of technology and specifically artificial intelligence in diagnosing patients provides a promising aiding tool in delivering correct and complete information to physicians. Previous research shows promising results for machine learning classifiers in recognizing and predicting a few frequent diseases.

Liang et al. (2019)[8] trained a multiclass linear logistic regression classifier on 101.6 million data points from 1,362,559 patients to predict 13 common diseases. The classifier had an average F1 score of 0.885 across these diseases. Similar to this study, multiple studies have examined ICD-9-CM code prediction. Baumel et al. (2018)[9] predicted ICD-9-CM codes using the unstructured portion of the MIMIC-III dataset, such as patient notes and evaluated four different approaches to classification and found that their Hierarchical Attention mechanism was the most successful with an F1 score of 55.86%. Lita et al. (2008)[10] predicts the five most frequent ICD-9-CM codes using multiple documents representing doctor's notes and test results for each patient. Lita et al. found an F1 score ranging from 0.568-0.729 for Support Vector Machine classifier and 0.6909-0.772 for Bayesian Ridge Regression.

While such early results on limited ranges of conditions are encouraging, one should bear in mind that physicians may have to consider thousands of possible diagnoses while the evaluated machine learning approaches were given the much easier task of selecting among fewer than 20 ICD codes. This study aims to address this disparity and evaluate machine learning models on growing datasets of increasingly infrequent diagnoses. In doing so, we hope to give a fair and realistic account of diagnostic decision support system quality.

**Methods**

The experiment followed four fundamental steps: (1) extraction and processing of data, (2) training various classifiers, (3) evaluating the top three classifiers' performance, (4) establishing feature importance. The main research question is how well standard machine learning techniques can learn a diagnosis proxy as represented by ICD-9 codes and how diagnosis prevalence influences their efficacy.

*Extraction and Processing of Data:*

We utilized the Medical Information Mart for Intensive Care III (MIMIC-III) database that holds data about 61,532 intensive care unit stays in the Beth Israel Deaconess Medical Center in Boston, Massachusetts from June 2001 to October 2012. The database holds detailed information about each admission, such as demographic information, lab measurements, and all chart information. There were 58,138 unique admissions into the ICU in the MIMIC-III database. For each admission, we used the ICD-9-CM code given a priority rank of one as the primary diagnosis (see a discussion on the limiting implications of this assumption at the end of this article). This ICD-9-CM code served as the predictive target for our machine learning systems.

In the MIMIC-III database, using SQL queries, we counted the number of admissions that had each unique ICD-9-CM code as their primary diagnosis. In this way, we established a list of most common diseases. Then we extracted features to represent each admission. We considered basic demographics (such as age, gender, marital status, and, religion), lab results, and chart information. Lastly, we determined the presence of co-morbidity ICD-9-CM codes that were not ranked at position one. The resulting feature vectors contained 51 attributes per admission. Each feature was chosen based on frequency in the MIMC dataset. We only included lab results and chart events completed for 60% of patients in our dataset. If multiple of the same lab results were measured, we took the mean of this sample to determine a singular value for the patient. If multiple chart events were recorded, we took the mode of this sample since chart events were categorical data. Missing values were imputed using the population mean of the respective attribute. Following common practice[11], we excluded any patients below the age of 18, leaving us with 54,717 total admissions. To study the effect of disease observation frequency on machine learning efficacy, we created five datasets containing the admissions with the top n target diagnoses (Table 1). Datasets with higher choices of n are supersets of all lower choices n. For example, the dataset reflecting all admissions corresponding to the top 100 diagnoses contains all admissions from the top 50 diagnoses. Each dataset is split into stratified training (80%) and test sets (20%).

**Table 1.** Number of ICD-9-CM codes, number of admissions in the dataset.

| Number of ICD-9-CM Codes: | Number of Admissions: |
|---|---|
| 50 | 27,615 |
| 100 | 34,720 |
| 200 | 40,293 |
| 500 | 47,031 |
| 1,000 | 50,905 |

*Model Comparison:*

Using the most common 50 disease dataset, we explored a wide range of different classifiers and hyper parameter settings using the Python scikit-learn library[12]. We considered logistic regression, support vector classifiers (SVC), nu-support vector classifiers (NuSVC), nearest neighbor classifiers, decision trees, random forests and multi-layer perceptrons (MLP). For each type of classifier, we ran the model on the default settings on scikit learn first. Then each different parameter was altered and optimized based on the results of the different runs. After multiple parameters were optimized, and the classifier was outputting similar F1 scores, we stopped altering parameters. A complete list of all 43 runs with the altered parameters is cited in the Appendix. The parameters listed were the ones altered. If the parameters are not listed, then the default from SciKitLearn was used.

*Prevalence vs. Efficacy:*

The best-performing models were trained and evaluated on each of the five prevalence-ranked datasets. As the datasets grew increasingly large, fewer patients were diagnosed with each individual primary diagnosis, giving machine models less diagnosis-specific training data.

*Feature Importance:*

We took the consistently most successful classifier, the MLP, and trained and evaluated it using the 50 disease dataset. The 50 disease dataset was picked for reasons of speed and computational efficiency. To test the importance of each feature, we performed an ablation study in which each of the 51 features was removed manually, and the MLP was trained on the remaining 50 features. The importance of each feature was then quantified by the change of the average F1 score of the run without that feature. After all 51 runs were completed, the features were ranked according to the difference in F1 score from the baseline (the model including all the features). We ran the same experiment but removed entire categories of features. We removed demographics, lab events, chart events, and co-morbidity and quantified the importance of each of those categories by the change in F1 score. A negative difference would mean that the absence of the corresponding feature hurt the model, suggesting that the feature helps predicting diagnoses. Conversely, a positive difference would mean that the feature was not indicative of the considered diagnoses and may even mislead the machine learning model.

**Results**

Our results are reported based on our three experiments: model comparison, varying diagnosis prevalence and studying feature importance.

*Model Comparison:*
Table 2 presents the ten most highly performing machine learning models together with their respective F1 scores on the held-out test set. Each classifier is followed by its location in the Appendix, which holds the specific parameters of that run. MLPs and Random Forests consistently achieved the highest scores followed by NuSVCs and other methods. While the top five methods' scores are closely tied, there is a noticeable drop in performance separating this leading group from the next best approaches.

**Table 2: Type of Classifier (number in appendix), Average F1 Score**

| Type of Classifier: | Average F1 Score: |
|---|---|
| MLP (1) | 0.281 |
| MLP (2) | 0.280 |
| MLP (3) | 0.282 |
| Random Forest (4) | 0.279 |
| MLP (5) | 0.278 |
| Random Forest (6) | 0.269 |
| NuSVC (7) | 0.260 |
| NuSVC (8) | 0.258 |
| NuSVC (9) | 0.255 |
| NuSVC (10) | 0.255 |

*Prevalence vs. Efficacy:*

Figure 1 plots the top three machine learning models' F1 scores as a function of the number of target diagnoses to be predicted. As hypothesized, the F1 scores progressively decrease as the number of examined ICD-9-CM codes increases. When predicting a broader range of diagnoses of individually lower prevalence the classification performance drops markedly.

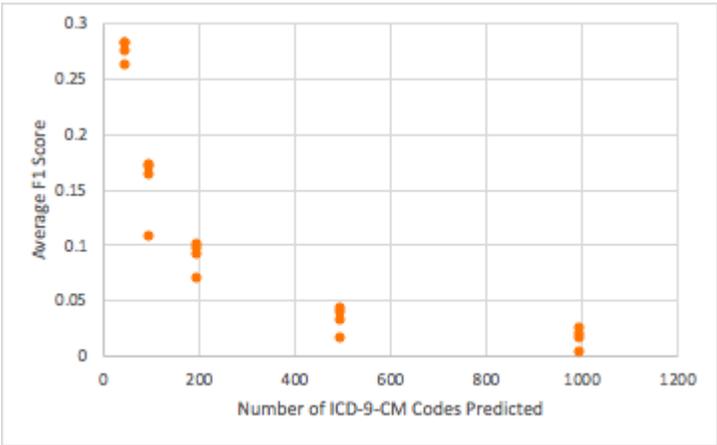

**Figure 1:** Distribution of average F1 scores over number of ICD-9-CM codes predicted.

To further demonstrate the effect of diagnosis prevalence independent of the number of classes to be predicted, Figure 2 displays the F1 score for predicting eachz individual ICD-9-CM code conditioned on the number of training data points for that specific diagnoses. To avoid clutter, we display the top two classifiers (MLP and Random Forest) and the 100 disease training dataset.

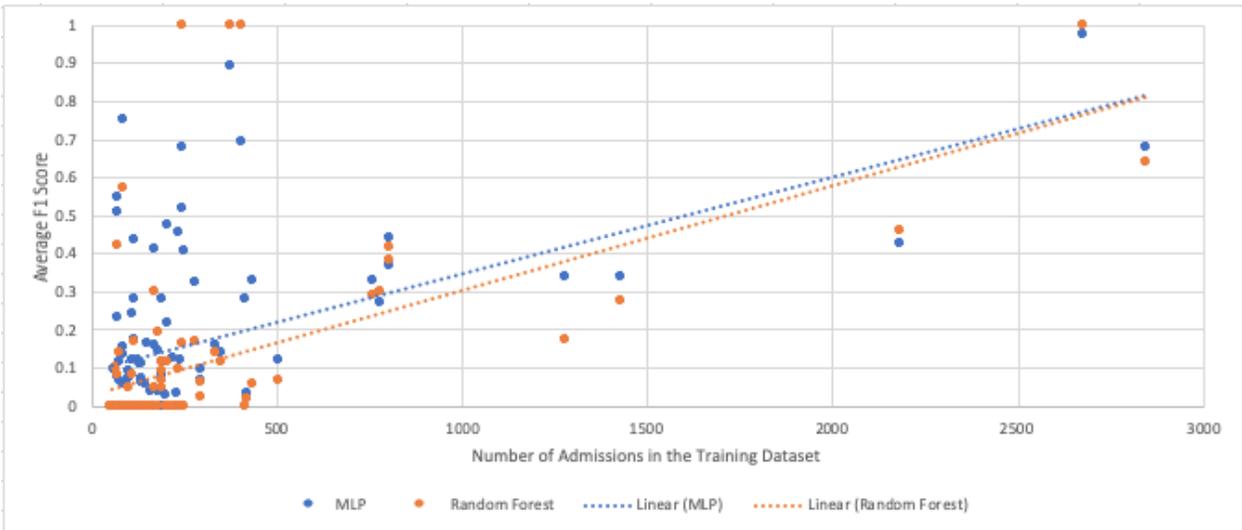

**Figure 2:** Distribution of average F1 score over number of entries (admissions) in the training dataset

The plot shows that diagnosis-specific F1 scores increase with increasing numbers of related admissions in the training dataset. To further test the relationship between F1 scores and the number of available observations, we calculated the Pearson correlation coefficient between F1 scores and the number of corresponding admissions in the training dataset. Table 3 lists the results for both models.

Table 3: Number of Diseases Examined, Average F1 Score, Correlation Coefficient, Random Performance

| Number of Diseases Examined: | MLP Avg F1 Score: | MLP Correlation Coefficient: | Random Forest Avg F1 Score: | Random Forest Correlation Coefficient: | Random Performance (1/n) |
|---|---|---|---|---|---|
| 50 | 0.28 | 0.5953 | 0.27 | 0.5833 | 1/50 |
| 100 | 0.17 | 0.5695 | 0.11 | 0.6153 | 1/100 |
| 200 | 0.09 | 0.60984302 | 0.09 | 0.60932512 | 1/200 |
| 500 | 0.04 | 0.64355 | 0.03 | 0.6084 | 1/500 |
| 1000 | 0.03 | 0.7033 | 0.01 | 0.6525 | 1/1000 |

The resulting correlation coefficients range from 0.5833 (a moderate positive relationship) to 0.70 (a strong positive relationship), indicating a systematic performance bias favoring diagnoses with higher numbers of available observations.

*Feature Importance:*

To better understand the previously presented machine learning models, Table 4 displays the change in F1 score with each feature being removed. For reasons of space, we only perform this ablation study for the single best performing MLP model. The three most important features for the MLP classifier were co-morbidity ICD-9-CM codes that the patient was diagnosed with. Additionally, different lab measurements such as heart rate and lymphocytes were important. Interestingly, marital status turned out to be an influential feature, demonstrating the potential for chance patterns or hidden covariates (i.e., marital status might be a proxy for age or economic status) in such patient databases.

Table 4: Feature Removed, Average F1 score of that run, Change in F1 score (Average F1 score without the feature subtracted by the average F1 score with all the features)

| Feature Removed: | F1 Score: | Change in F1 Score: | Feature Removed: | F1 Score: | Change in F1 Score: |
|---|---|---|---|---|---|
| ICD 9 code 414.01 (Coronay atherosclerosis of native coronary artery) | 0.239963663 | -0.040225611 | magnesium | 0.277745208 | -0.002444066 |
| ICD9 code 427.31 (Atrial fibrillation) | 0.248887276 | -0.031301998 | bilirubin totall | 0.277842942 | -0.002346332 |
| ICD 9 code 584.9 (Acute kidney failure) | 0.258504003 | -0.021685271 | chloride | 0.278009066 | -0.002180208 |
| lymphocytes | 0.2647802 | -0.015409074 | ph | 0.278345652 | -0.001843622 |
| marital status | 0.265188897 | -0.015000377 | protein | 0.278472925 | -0.001716349 |
| heart rate | 0.266713622 | -0.013475652 | base excess | 0.278481555 | -0.001707719 |
| ptt | 0.267462665 | -0.012726609 | specific gravity | 0.278704116 | -0.001485158 |

Table 4 (Continued):

| Feature Removed: | F1 Score: | Change in F1 Score: | Feature Removed: | F1 Score: | Change in F1 Score: |
|---|---|---|---|---|---|
| po2 | 0.26941512 | -0.010774154 | pt | 0.278982071 | -0.001207203 |
| activity | 0.269844165 | -0.010345109 | urea nitogren | 0.279109843 | -0.001079431 |
| hemoglobin | 0.269890861 | -0.010298413 | whiteblood | 0.27945553 | -0.000733744 |
| mch | 0.270363972 | -0.009825302 | hematocrit | 0.280174599 | -1.46745E-05 |
| red blood | 0.270474954 | -0.00971432 | none | 0.280189274 | 0 |
| ICD9 code 428.0(Congestive heart failure) | 0.270569877 | -0.009619397 | gender | 0.28035032 | 0.000161046 |
| monocytes | 0.271803736 | -0.008385538 | sodium | 0.280653758 | 0.000464484 |
| pco2 | 0.27312788 | -0.007061394 | ICD 9 code 401.9 (Hypertension) | 0.280873915 | 0.000684641 |
| calcium(total) | 0.273298219 | -0.006891055 | platet count | 0.281675965 | 0.001486691 |
| rdw | 0.273508764 | -0.00668051 | neutrophils | 0.281813962 | 0.001624688 |
| potassium | 0.273770778 | -0.006418496 | mcv | 0.282209289 | 0.002020015 |
| glucose | 0.275423143 | -0.004766131 | mchc | 0.284459708 | 0.004270434 |
| ph2 | 0.27561567 | -0.004573604 | age | 0.284511244 | 0.00432197 |
| bowel sounds | 0.275696288 | -0.004492986 | phosphate | 0.285002243 | 0.004812969 |
| inr(pt) | 0.276359803 | -0.003829471 | calculate total co2 | 0.287402328 | 0.007213054 |
| eosinophils | 0.276446644 | -0.00374263 | basophils | 0.289307738 | 0.009118464 |
| anion gap | 0.276858886 | -0.003330388 | religion | 0.290350027 | 0.010160753 |
| creatinine | 0.276859854 | -0.00332942 | lactate | 0.293504219 | 0.013314945 |
| bicarbonate | 0.277162144 | -0.00302713 | | | |

Table 5 displays the results of the identical experiment but with the aggregate features removed ranked by importance or largest drop in F1 score. Unsurprisingly, since lab events were the majority of features collected (74%) the removal of that data resulted in the sharpest decrease in F1 score of the model. Similarly, co-morbidity was the second largest category of features collected and therefore resulted in the second largest drop in F1 score.

**Table 5.** Feature Removed, Average F1 score of that run, Change in F1 score (Average F1 score without the feature subtracted by the average F1 score with all the features)

| Feature Removed | Average F1 Score | Change in F1 Score |
| --- | --- | --- |
| Lab Events | 0.161795 | 0.118393945 |
| Co-Morbidity | 0.208749 | 0.071440546 |
| Chart Events | 0.272903 | 0.007286595 |
| Demographics | 0.278865575507839 | 0.001323698 |

**Discussion**:

The overall goal of this study was to establish the relationship between the prevalence of a diagnosis in a patient cohort and the performance of machine learning models trained on that cohort. Our analysis indicates that the average F1 score drops as the number of admissions with that disease decreases as well. However, we made three further observations.

First, overall the current classifiers are not sufficiently effective in predicting the vast majority of ICD-9-CM codes with this dataset. For example, the best performing classifier had an average F1 score of 0.282 when trying to predict the 50 most common ICD-9-CM codes. Even though these ICD-9-CM codes are fairly common, all classifiers still performed poorly. In the top performing classifier (MLP) only one ICD-9-CM code had an F1 score above 0.90 and only six of the ICD-9-CM codes had F1 scores above 0.50. These ICD-9-CM codes, as predicted, were those with the highest number of admissions in the training dataset. Additionally, the six ICD-9-CM codes with the lowest numbers of corresponding admissions in the 50 ICD-9-CM code training set received an F1 score of 0.0. In the context of diagnostic decision support, error rate tolerance is very low. Depending on disease severity and the resulting cost of misses and false alarms, an error rate of 1 percent may still be too high.

Second, to avoid missing values, we selected lab and chart events via observation frequency. That is, we included only those vital signs that were measured for most patients in our database. Interestingly, this criterion does not imply feature importance. The five most frequently observed vital signs (hematocrit, white blood cells, platelet count, mchc (mean corpuscular hemoglobin concentration) and red blood) were measured for 97.88% of admissions but each resulted in negligible performance score drops when removed. Additionally, the lab measurement of lymphocytes and po2 (partial pressure of oxygen) proved to be significant in helping the classifier despite only having been ordered for 43,347 (74.5%) and 37,267 (64.10%) of all admissions, respectively. This illustrates that while completeness of observation is a desirable property of features, it does not guarantee predictive power.

Third, the features with the greatest observed importance did not always correspond to clinical intuition. We initially expected common vital signs and lab measurements to aid the classifier the most, as they are general-purpose descriptors of patient state. However, none of the top five features match that description. Other demographic information, such as gender and religion turned out to be insignificant in terms of predictive power.

This study additionally had limitations. First, we assumed that the top-ranked ICD-9-CM code reflected the correct diagnosis. This assumption was necessary as no other structured data was available. However, it has been frequently pointed out that ICD-9-CM codes are assigned mainly for billing purposes and might not always accurately reflect diagnoses[13]. Second, selecting features based on observation frequency loses potential information about a patient's diagnosis since for example a lab exam would not be missing at random. Therefore, the lab tests doctors order are correlated with the diagnosis a doctor believes a patient to have. Third, mean imputation carries a heavy assumption that data are missing completely at random. Since discussed above how the missing data is not random, using this method to impute data will likely introduce bias into the analysis. Fourth, the MIMIC dataset utilizes data from the ICU. However, an ICU database is an "extreme" situation where ICU patients often have highly complex multi-morbidity. Therefore, these complex diagnoses increases the difficulty of ML classifiers to accurately predict the diagnosis.

Future work and next steps will consider organizing groups of diagnoses into categories, such as defined by CCS[14]. and using those categories (rather than individual co-morbidity ICD codes) as features. Additionally, we will draw from recent advances in transparent and explainable machine learning methods to better understand feature importance. Finally, we will investigate the use of robust machine learning techniques to better scale to scenarios involving large numbers of individually infrequent diagnoses.

**Conclusion**

The overwhelming majority of machine learning studies for diagnostic decision support consider only a limited number of highly prevalent conditions to distinguish between. While recently proposed models such as deep neural networks have become adept at solving this task, there is a marked difference from the diagnostic scenario that most physicians face in reality. Here, the range of diagnoses is not as clearly delineated and automatic models that can only consider highly frequent conditions such as congestive heart failure or diabetes may be of limited use.

This study investigated the relationship between diagnosis prevalence and machine learning model performance, showing that, indeed, standard models perform well for the few most frequent conditions and rapidly deteriorate in performance as the range of considered conditions grows to more realistic scopes.

**Appendix**

Table 6. Rank, Type of Classifier, Hyper-Parameters Altered, F1 Score

| Rank | Type of Classifier | Hyper-Parameters Altered | F1 Score |
|---|---|---|---|
| 1 | MLP | Activation function= logistic<br>Batch size= 500<br>Learning rate= adaptive | 0.2818 |
| 2 | MLP | Activation function= logistic<br>Learning rate= adaptive | 0.2800 |
| 3 | MLP | Activation function= logistic | 0.2822 |
| 4 | Random Forest | Number of weak learners= 200<br>Maximum features= 49 | 0.2786 |
| 5 | MLP | None | 0.2784 |
| 6 | Random Forest | Number of weak learners = 200 | 0.2690 |
| 7 | NuSVC | nu= 0.08<br>gamma= scale<br>shrinking= false | 0.2606 |
| 8 | NuSVC | nu=0.081<br>gamma=scale | 0.2575 |
| 9 | NuSVC | nu=0.081<br>gamma=scale | 0.2554 |
| 10 | NuSVC | nu=0.08<br>gamma=scale<br>decision function shape= ovo | 0.2554 |
| 11 | NuSVC | nu=0.05<br>gamma=scale | 0.2356 |
| 12 | MLP | Activation function= tanh | 0.2214 |
| 13 | Random Forest | Number of weak learners =200<br>minimum samples split=20<br>minimum samples leaf=10 | 0.2209 |
| 14 | Decision Tree | Minimum samples split=20,<br>minimum samples leaf=10,<br>minimum impurity decrease=0.0001 | 0.198600946 |
| 15 | Decision Tree | Maximum features=49,<br>minimum samples split=10,<br>minimum samples leaf=10 | 0.198294587 |
| 16 | Decision Tree | Maximum depth= 100 | 0.193617647 |
| 17 | Random Forest | Number of weak learners =200maximum features=49<br>bootstrap=False | 0.188815205 |
| 18 | Decision Tree | Minimum samples split=50,<br>minimum samples leaf=50 | 0.188160476 |
| 19 | Decision Tree | Maximum depth=15 | 0.185721399 |
| 20 | Decision Tree | Maximum depth=50 | 0.185533647 |
| 21 | NuSVC | nu=0.01, gamma=scale | 0.185040159 |
| 22 | Decision Tree | Maximum depth= 30 | 0.181199121 |
| 23 | Decision Tree | none | 0.177232993 |
| 24 | Decision Tree | splitter=random maximum depth=100 | 0.160753496 |
| 25 | Logistic Regression | None | 0.128222532 |
| 26 | KNeighbor Classifier | Number of neighbors=1 | 0.121501548 |
| 27 | KNeighbor Classifier | Number of neighbors=1 weights=distance | 0.121501548 |
| 28 | KNeighbor Classifier | None | 0.120030598 |
| 29 | KNeighbor Classifier | Number of neighbors= 20 | 0.112270134 |
| 30 | MLP | activation=logistic solver=lbfgs | 0.102143673 |
| 31 | KNeighbor Classifier | Number of neighbors= 50 | 0.101861607 |
| 32 | Radius Neighbor Classifier | radius=50 weights=distance | 0.099851443 |
| 33 | Radius Neighbor Classifier | radius=50 | 0.097489706 |
| 34 | SVC | gamma=scale | 0.073679536 |
| 35 | SVC | Decision function shape= ovo gamma=scale | 0.073679536 |
| 36 | Radius Neighbor Classifier | Radius=75 | 0.067067629 |
| 37 | Radius Neighbor Classifier | Radius=20 | 0.031960733 |
| 38 | Logistic Regression | none | 0.028452085 |
| 39 | MLP | solver= lbfgs | 0.02785047 |
| 40 | MLP | solver= sgd | 0.025819759 |
| 41 | SVC | none | 0.017584537 |
| 42 | Radius Neighbor Classifier | none | 0 |
| 43 | Radius Neighbor Classifier | radius = 1 | 0 |